\pdfoutput=1

\documentclass[aps,prl,preprintnumbers,showkeys,twocolumn,amsmath,amssymb,longbibliography,floatfix]{revtex4-1}%

\usepackage[pdftex,colorlinks,hyperindex,plainpages=false,bookmarksopen,bookmarksnumbered,pdfusetitle,allcolors={blue}]{hyperref}

\usepackage{amsmath}
\usepackage{bm}
\usepackage{graphicx}
\usepackage{float}
\usepackage{mathrsfs}
\usepackage{bm}
\usepackage{bbold}
\usepackage{cases}
\usepackage{comment}

\newcommand\dd{\text{d}}

\usepackage{tikz}
\begin{document}

\title{Electrically Controlled Crossed Andreev Reflection in Two-Dimensional Antiferromagnets}

\author{Martin F.~Jakobsen}
\affiliation{Center for Quantum Spintronics, Department of Physics, Norwegian University of Science and Technology, NO-7491 Trondheim, Norway}
\author{Arne Brataas}
\affiliation{Center for Quantum Spintronics, Department of Physics, Norwegian University of Science and Technology, NO-7491 Trondheim, Norway}
\author{Alireza Qaiumzadeh}
\affiliation{Center for Quantum Spintronics, Department of Physics, Norwegian University of Science and Technology, NO-7491 Trondheim, Norway}

%%%%%%%%%%%%%%%%%%%%%%%%%%%%%%%%%%%%%%%%%%%%%%%%%%%
\begin{abstract}
We report generic and tunable crossed Andreev reflection (CAR) in a superconductor sandwiched between two antiferromagnetic layers. We consider recent examples of two-dimensional magnets with hexagonal lattices, where gate voltages control the carrier type and density, and predict a robust signature of perfect CAR in the nonlocal differential conductance with one electron-doped and one hole-doped antiferromagnetic lead. The magnetic field-free and spin-degenerate CAR signal is electrically controlled and visible over a large voltage range, showing promise for solid-state quantum entanglement applications.
\end{abstract}
\maketitle

%%%%%%%%%%%%%%%%%%%%%%%%%%%%%%%%%%%%%%%%%%%%%%%%%%%

%INTRODUCTION
\textit{Introduction}.-- In quantum mechanics, identical particles can form entangled pairs sharing a common wave function: a measurement performed on one particle predetermines the state of the other. Entanglement is a unique quantum effect and was first experimentally verified using pairs of linearly polarized photons \cite{PhysRevLett.49.1804,Ursin2007}. Currently, entangled states play a vital role in quantum computing, communication, and cryptography technologies \cite{PhysRevA.57.120,RevModPhys.81.865}. Nevertheless, large-scale societal implementation requires entanglement in solid-state devices over long distances.

Electrons in a Cooper pair can be spatially separated, but remain spin and momentum entangled via a process called Cooper pair splitting \cite{pandey2020ballistic, fuchs2020crossed, Zhang_2018}. The time-reversed process is called crossed Andreev reflection (CAR) or nonlocal Andreev reflection. CAR is the nonlocal process of converting an incoming electron from one voltage-biased lead into an outgoing hole in another grounded lead via Cooper pair formation in a grounded superconductor~\cite{PhysRevLett.74.306,doi:10.1063/1.125796}. This process requires the distance between the two leads to be comparable to or shorter than the superconducting coherence length. A significant disadvantage in current state-of-the-art technology is that two detrimental processes often mask the CAR signal: (i) nonlocal electron cotunneling (CT) between the two leads and (ii) local Andreev reflection (AR) in the voltage-biased lead.  The optimal solution, is to design a system that suppresses CT and AR signals while enhancing CAR signals.

Presently, numerous superconducting heterostructures have been proposed to enhance CAR signals utilizing different leads, such as normal metals (NMs) \cite{Falci_2001,PhysRevB.75.172503,PhysRevB.78.224515,PhysRevB.93.214508}, ferromagnetic (FM) metals \cite{doi:10.1063/1.125796,PhysRevB.80.014513,PhysRevB.80.014513,PhysRevB.94.165441,PhysRevB.96.161403, PhysRevLett.97.087001}, two-dimensional (2D) graphene \cite{PhysRevLett.100.147001,PhysRevLett.105.107002}, and topological insulators \cite{PhysRevLett.101.120403,PhysRevLett.109.036802,PhysRevB.84.115420,PhysRevLett.110.226802,He2014,PhysRevB.91.085415}. Conclusive experimental detection of CAR signals remains challenging, but progress has been made by utilizing NM leads \cite{Kleine_2009, PhysRevLett.95.027002, PhysRevLett.97.237003, Cadden-Zimansky2009, Wei2010, Das2012}, FM leads \cite{PhysRevLett.93.197003,PhysRevB.74.092501,Beckmann2007,https://doi.org/10.1002/andp.200510154,PhysRevB.72.184515}, quantum dots \cite{PhysRevLett.114.096602,PhysRevLett.115.227003,Hofstetter2009,PhysRevLett.107.136801,PhysRevLett.109.157002, PhysRevLett.104.026801,Borzenets2016}, and very recently, graphene-based systems with opposite doping levels in the two leads \cite{pandey2020ballistic,Park2019}.
Nevertheless, most proposals require fine-tuning of the electronic structure and bias voltage. Furthermore, there are two additional disadvantages associated with FM leads: First, stray fields limit the potential use of FM systems in high-density applications. Second, although in FM half metal leads, it is possible to enhance the CAR signal when the magnetization of two leads is antiparallel, the spin entanglement of the electrons is simultaneously lost ~\cite{PhysRevLett.105.107002,PhysRevB.80.014513}.

In this Letter, we propose utilizing 2D metallic antiferromagnetic leads, to overcome these issues. Although antiferromagnetic systems are magnetically staggered ordered systems, they have negligible stray fields and their degenerate spin states preserve entanglement.

Recently, antiferromagnets have revealed potential in superconducting spintronics.
For instance, at the antiferromagnet-superconductor interface, normal electron reflection (NR) and AR have been demonstrated to be both specular and retroreflective~\cite{PhysRevB.102.140504,PhysRevLett.94.037005,PhysRevLett.96.117005,Zhou_2019,PhysRevB.95.104513,PhysRevB.88.214512,PhysRevB.72.184510,PhysRevResearch.1.033095,PhysRevB.68.144517,PhysRevLett.99.017004,H_bener_2002,Constantinian2013,doi:10.1063/1.4824891, PhysRev.142.118,seeger2021penetration}. In heterostructures, these anomalous processes fundamentally change the behavior of the electrical and thermal conductance \cite{PhysRevB.102.140504}. In Josephson junctions, atomic-scale $0$-$\pi$ transitions~\cite{PhysRevLett.96.117005,Zhou_2019,PhysRevB.95.104513,PhysRevB.88.214512,PhysRevB.72.184510,PhysRevResearch.1.033095} are predicted to occur. The existence of Josephson effects has been experimentally verified~\cite{PhysRevB.68.144517,PhysRevLett.99.017004,H_bener_2002,Constantinian2013,doi:10.1063/1.4824891,PhysRev.142.118,seeger2021penetration}, but the remaining theoretical predictions have yet to be explored.

Herein, we investigate the suitability of an antiferromagnet-superconductor-antiferromagnet (AF-S-AF) junction with a 2D hexagonal lattice as a platform for experimental detection and quantum applications of CAR signals. Our model is general and applicable to systems in which antiferromagnetism and superconductivity are either intrinsic to the material or induced by proximity. We find a gate-controllable window in parameter space, wherein both CT and AR signals can be completely suppressed in favor of the CAR signal. This robust experimental signature is expected to be directly measurable over a wide range of applied bias voltages. Our prediction of enhanced CAR signals in antiferromagnetic-based devices combined with recent experimental advances in graphene-based junctions \cite{pandey2020ballistic,Park2019} open a unique opportunity to realize efficient large-scale Cooper pair splitters with immediate applications in solid-state quantum entanglement technology.

\begin{figure}[t]
\centering
\includegraphics[width=\columnwidth]{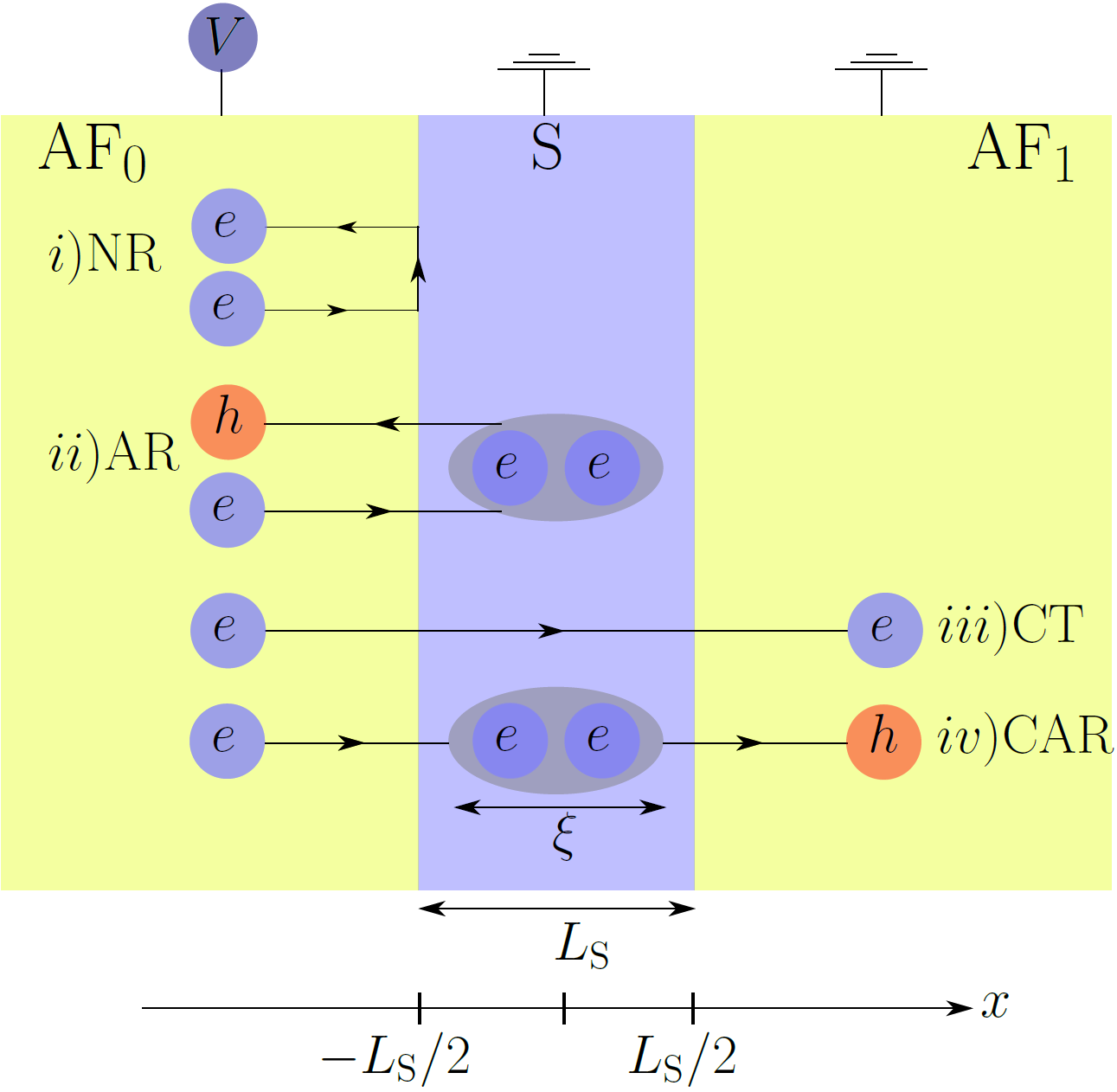}
\caption{The scattering processes in the AF-S-AF junction. We assume that $\mathrm{AF}_0$ is biased with voltage $V$, while $\mathrm{S}$ and $\mathrm{AF}_1$ are grounded. An incoming electron in $\mathrm{AF}_0$ may undergo (i) NR, (ii) AR, (iii) CT, or (iv) CAR. NR and AR contribute to the local conductance measured in $\mathrm{AF}_0$, and CT and CAR contribute to the nonlocal conductance measured in $\mathrm{AF}_1$.}
\label{fig:system}
\end{figure}

%%%%%%%%%%%%%%%%%%%%%%%%%%%%%%%%%%%%%%%%%%%%%%%%%%%
%MODEL AND THEORY
\textit{Model.}-- We consider a superconductor of length $L_{\mathrm{S}}$ sandwiched between two semi-infinite 2D antiferromagnetic metals with hexagonal lattices, forming a 2D AF-S-AF junction along the $x$ direction, as shown in Fig. \ref{fig:system}. The left lead $(\mathrm{AF}_0)$, the superconductor (S), and the right lead $(\mathrm{AF}_1)$ occupy the regions $x<-L_{\mathrm{S}}/2$, $-L_{\mathrm{S}}/2<x<L_{\mathrm{S}}/2$, and $x>L_{\mathrm{S}}/2$, respectively. The dynamics of charge carriers around the $K$ point in the Brillouin zone are governed by an eigenvalue problem $H(x) \Psi(x) = E \Psi(x)$, where
\begin{equation}
    H(x) =
    \begin{bmatrix}
    H^e_\mathrm{AF}(x)-E_\mathrm{F}(x) & \tilde{\Delta}(x) \\
    \tilde{\Delta}^{\dagger}(x) & H^h_\mathrm{AF}(x)+E_\mathrm{F}(x)
    \end{bmatrix}
    \label{Eq:DBDG}
\end{equation}
is the mean-field Bogoliubov de-Gennes (BdG) Hamiltonian \cite{PhysRevLett.97.067007,RevModPhys.80.1337}, and $E_{\mathrm{F}}(x)$ is the local Fermi energy. In 2D systems, $E_{\mathrm{F}}(x)$ may be tuned by a gate voltage.

\begin{figure}[t]
\centering
\includegraphics[width=\columnwidth]{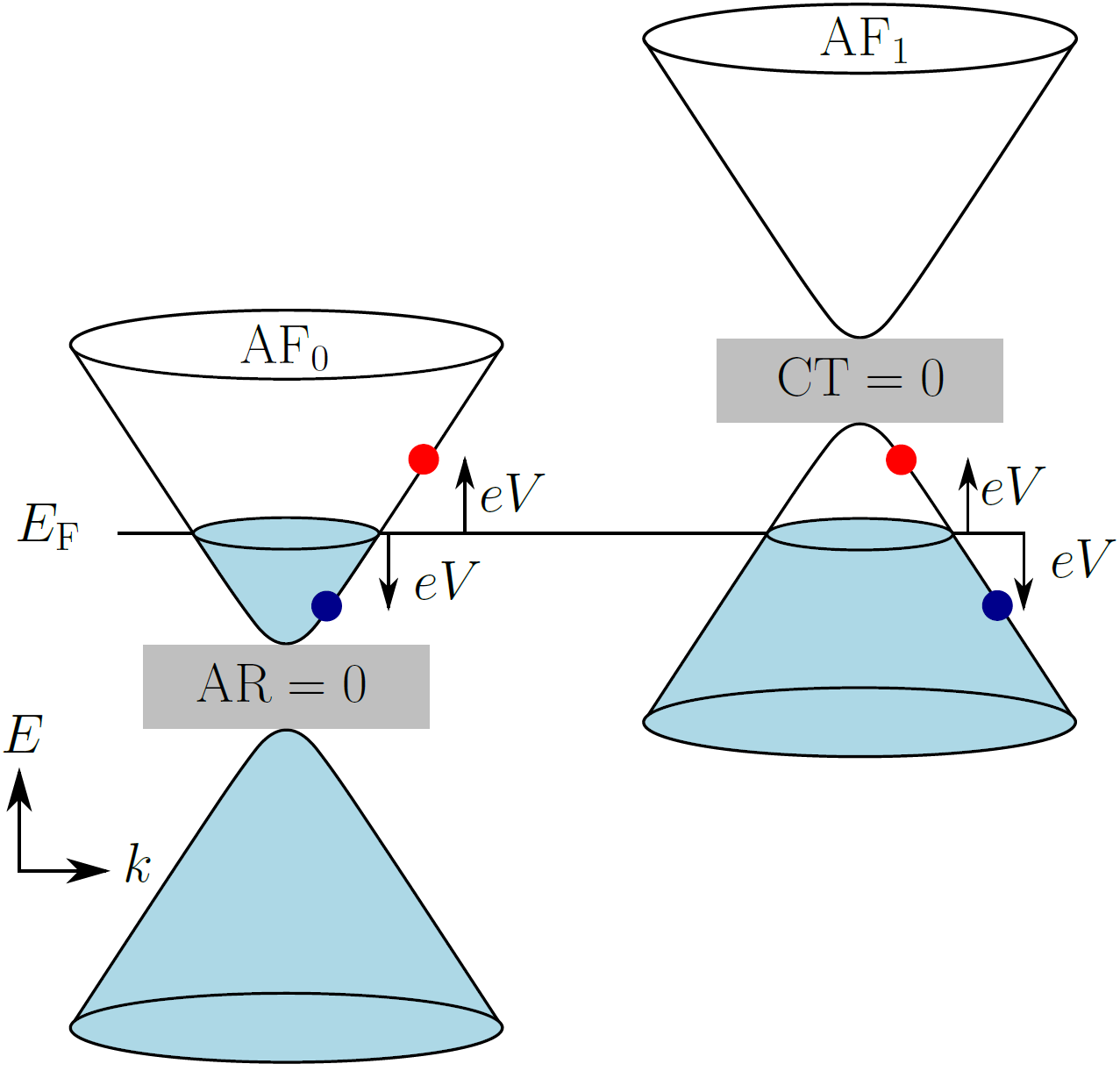}
\caption{The ''relativistic'' dispersions of itinerant electrons in 2D antiferromagnetic hexagonal lattices in the leads $\mathrm{AF}_0$ and $\mathrm{AF}_1$ are shown to the left and right, respectively. Electrons (holes) are denoted by red (blue) circles. It is possible to block both AR and CT signals to favor CAR signals by tuning the local Fermi energy close to the gap induced by the antiferromagnetic exchange interaction (gray region).}
\label{fig:Dispersion_AF}
\end{figure}

In the BdG Hamiltonian, the dynamics of low-energy itinerant charge carriers in the hexagonal antiferromagnetic leads around the $K$ point are described by an effective Dirac-like Hamiltonian for the electron subsector $H^e_\mathrm{AF}=H_{\bm{p}} + H_\mathrm{sd}$ and the hole subsector $H^h_\mathrm{AF}=-H_{\bm{p}} - H^T_\mathrm{sd}$, where $T$ denotes the transpose operator. The kinetic Hamiltonian of the antiferromagnet is
\begin{equation}
    H_{\bm{p}}(x)  =   v_{\mathrm{F}}s_0 \otimes \left(\bm{\sigma} \cdot \bm{p}\right),
\end{equation}
where $v_{\mathrm{F}}$, $\bm{p}=-i \hbar \left(\partial_x,\partial_y\right)$, and $\hbar$ denote the Fermi velocity, 2D momentum operator, and reduced Planck constant, respectively. In our notation, $\bm{\sigma}$ and $\bm{s}$ denote the Pauli matrices in sublattice and spin space, respectively.
The antiferromagnetic s-d exchange interaction between localized magnetic moments and itinerant electron spins
are described by
\begin{equation}
    H_\mathrm{sd}(x)=J(x)\left(\mathbf{n}(x)\cdot \mathbf{s}\right)\otimes \sigma_z.
\end{equation}
Here, $J(x)$ and $\bm{n}(x)$ denote the exchange strength and staggered N\'eel vector direction, respectively. We consider single-domain and collinear AFs with a uniform N\'eel vector in each lead, $\bm{n}(x) = \bm{n}_j$, where the index $j=\{0,1\}$ refers to the lead $\mathrm{AF}_j$. The misalignment angle between the N\'eel vectors is $\delta \gamma = \arccos{\left(\bm{n}_0 \cdot \bm{n}_1\right)}$. The eigenenergies of the 2D antiferromagnetic hexagonal Hamiltonian $H^{e(h)}_\mathrm{AF}$ are $E_\mathrm{AF}=\pm\sqrt{(\hbar v_F \bm{k})^2+J^2}$, where $\bm{k}$ is the 2D wave vector and $+$ and $-$ refer to the conduction and valence bands, respectively. Thus, the itinerant charge carriers around the $K$ point behave like massive Dirac particles with a band gap of magnitude $2J$ induced by the antiferromagnetic \textit{s-d} exchange interaction (see Fig. \ref{fig:Dispersion_AF}).

We consider an \textit{s}-wave superconductor described by BCS theory where the superconducting gap in the two-sublattice space is
\begin{equation}
    \tilde{\Delta}(x) = i s_y \otimes \Delta(x)\sigma_0.
\end{equation}
The superconducting coherence length is given by $\xi = \hbar v_{\mathrm{F}}/\Delta$, which estimates the Cooper pair size. The mean-field requirement of superconductivity is that the local Fermi energy in the superconductor, $E_{\mathrm{FS}}$, is much larger than the superconducting gap.

To illustrate the main concepts, and for clarity and simplicity, we assume that all energy scales exhibit the step-function behavior:
\begin{equation}
   \left\{J(x),E_{\mathrm{F}}(x),\Delta(x)\right\}=
    \left\{
        \begin{array}{ll}
         \left\{J_0,E_{\mathrm{F}0},0\right\}, & x< - \frac{L_{\mathrm{S}}}{2}, \\
          \left\{0,E_{\mathrm{FS}},\Delta_0\right\}, & - \frac{L_{\mathrm{S}}}{2} < x < \frac{L_{\mathrm{S}}}{2}, \\
         \left\{J_1,E_{\mathrm{F}1},0\right\}, & x>  \frac{L_{\mathrm{S}}}{2},
        \end{array}
    \right.
\end{equation}
where $\{J_j,E_{\mathrm{F}j}, E_{\mathrm{FS}},\Delta_0\}$ are constants and $j = \{0,1\}$ refers to the lead $\mathrm{AF}_j$. We also assume that $E_{\mathrm{FS}} \gg E_{\mathrm{F}j}$, and that the interfaces are magnetically compensated and ideal. Interface effects are discussed in the Supplemental Material (SM)~\cite{supplement}.

\textit{Local and nonlocal conductance.}-- We consider that $\mathrm{AF}_0$ is biased with voltage $V$, while S and $\mathrm{AF}_1$ are grounded. Consequently, a local and nonlocal conductance can be measured in $\mathrm{AF}_0$ and $\mathrm{AF}_1$, respectively. To determine the local and nonlocal conductance, we consider a scattering problem with an incident electron from $\mathrm{AF}_0$. In general, the allowed scattering processes are (i) local NR, (ii) local AR, (iii) nonlocal CT, and (iv) nonlocal CAR, as shown schematically in Fig. \ref{fig:system}. Using the Blonder-Tinkham-Klapwijk formalism \cite{PhysRevB.25.4515}, the local conductance
\begin{equation} \label{G_l}
    G_{\mathrm{L}} = \sum_{s=\uparrow,\downarrow}\int_{-\infty}^{\infty} \dd \varepsilon \left(-\frac{\partial f}{\partial \varepsilon}\right)  G_0^s\left(2 - G^s_{\mathrm{NR}} + G^s_{\mathrm{AR}}\right)
\end{equation}
is determined by NR and AR, while the nonlocal conductance
\begin{equation} \label{G_nl}
    G_{\mathrm{NL}} = \sum_{s=\uparrow,\downarrow}\int_{-\infty}^{\infty} \dd \varepsilon \left(-\frac{\partial f}{\partial \varepsilon}\right) G_1^s \left(G^s_{\mathrm{CT}} - G^s_{\mathrm{CAR}}\right)
\end{equation}
is determined by CT and CAR. Note that CT and CAR contribute with opposite signs in Eq. (\ref{G_nl}). Herein, $s$ denotes the spin degree of freedom, and $G^s_j$ is the intrinsic conductance of the lead $\mathrm{AF}_j$. The Fermi-Dirac distribution of incident electrons in $\mathrm{AF}_0$ at temperature $T$ is denoted by $f = \left(e^{\beta(\varepsilon-eV)} + 1\right)^{-1}$, where $\beta$ is the thermodynamic beta and $e$ is the elementary charge. Explicitly,
\begin{equation}
\begin{split}
    &G_{\mathrm{NR}(\mathrm{AR})}^s = \int_{-\pi/2}^{\pi/2} \dd \theta\, \cos \theta\,  R_{e(h)}^s,\\
    &G_{\mathrm{CT}(\mathrm{CAR})}^s = \int_{-\pi/2}^{\pi/2} \dd \theta\, \cos \theta\,  T_{e(h)}^s,
\end{split}
\end{equation}
where $R_{e(h)}^s$ and $T_{e(h)}^s$ are the spin-dependent probabilities of NR (AR) and CT (CAR), respectively, and $\theta$ is the angle of incidence for the incoming electron (see SM~\cite{supplement}). In the following, we consider the zero temperature limit.

%%%%%%%%%%%%%%%%%%%%%%%%%%%%%%%%%%%%%%%%%%
%RESULTS
\textit{CAR enhancement.}--
As mentioned above, the antiferromagnetic \textit{s-d} exchange interaction induces a band gap of $2J_j$ in each lead $\mathrm{AF}_j$. When a gate voltage is used to tune the local Fermi energy $E_{\mathrm{F}j}$, it is possible to control the contributions of different scattering processes to the total nonlocal conductance. As an example, consider the case in which $E_{\mathrm{F}0} = - E_{\mathrm{F}1} = E_{\mathrm{F}}>J_j >0$, where $\mathrm{AF}_0$ is electron doped and $\mathrm{AF}_1$ is hole doped, as depicted in Fig. \ref{fig:Dispersion_AF}. In this case, CT is completely suppressed for bias voltages in the interval $E_{\mathrm{F}} - J_{1} < eV < E_{\mathrm{F}} + J_{1}$. Furthermore, if we set $J_0 = J_1 = E_{\mathrm{F}}$ and $E_{\mathrm{F}}/\Delta_0> 1/2$, then the CAR signal becomes dominant for all voltages in the subgap regime, $eV/\Delta_0<1$. 

\begin{figure}[t]
\centering
\includegraphics[width=\columnwidth]{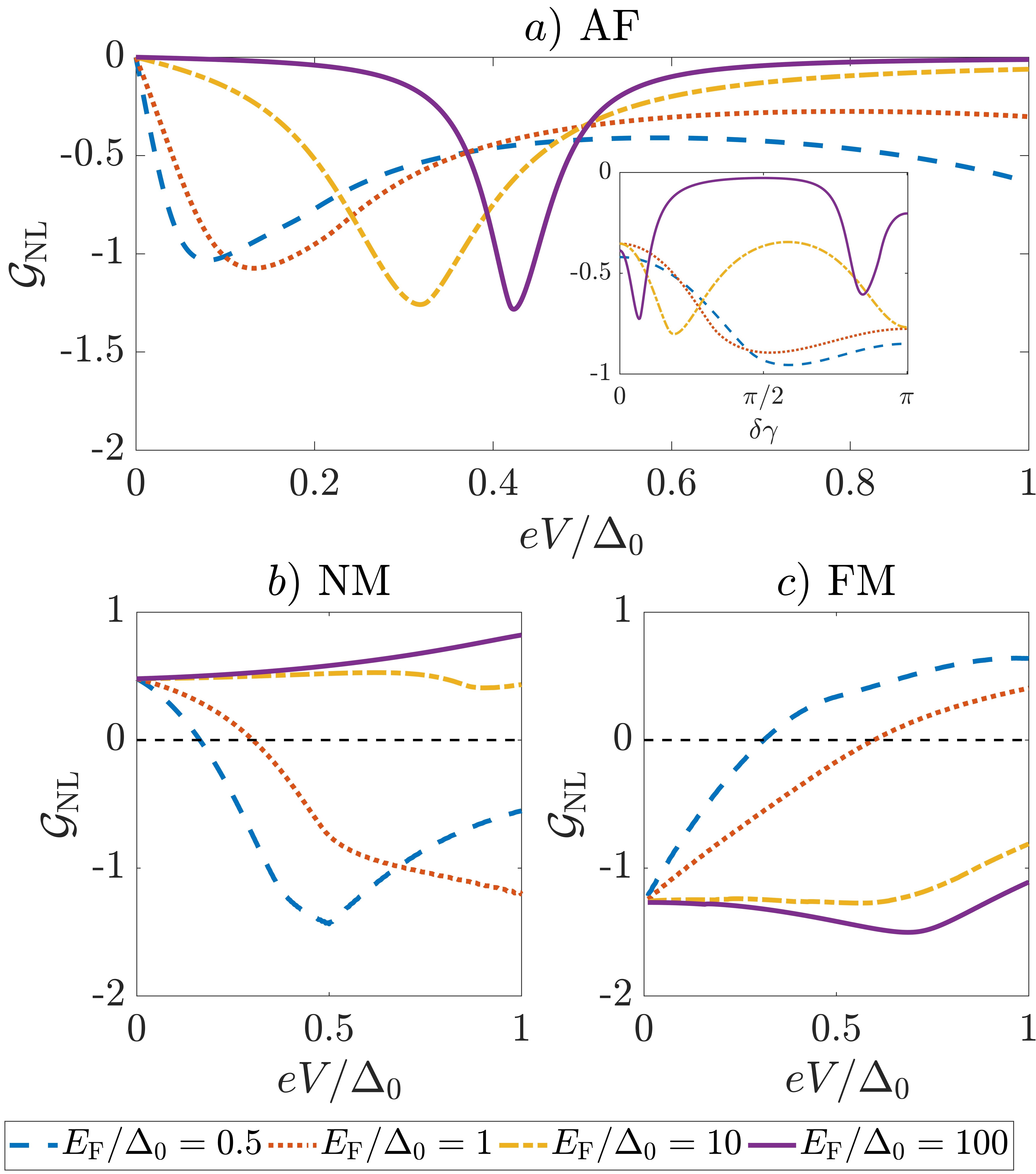}
\caption{The total nonlocal conductance in CAR-dominant configurations as a function of the applied voltage bias for different 2D hexagonal heterostructures. (a) An AF-S-AF system with parallel N\'eel vectors and opposite charge doping in the leads, (b) an NM-S-NM system with opposite charge doping in the leads, and (c) an FM-S-FM system with antiparallel magnetization vectors and the same charge doping in the leads.
The inset in (a) shows the angular dependence of the nonlocal conductance versus the misalignment between the N\'eel vectors in the two leads $\delta \gamma = \arccos{\left( \bm{n}_0\cdot \bm{n}_1\right)}$, with an applied voltage bias $eV/\Delta_0= 0.5$.
In all figures, we have set the AF (FM) \textit{s-d} exchange interaction equal to the Fermi energy in both leads.
}
\label{fig:NLGvsEV}
\end{figure}

To study the CAR-dominant regime, we set $J_0 = J_1 = J = E_{\mathrm{F}} > 0$, where both AR and CT processes are suppressed simultaneously. For concreteness, we fix the length of the superconductor to its coherence length $L_{\mathrm{S}} = \xi$ and assume that the N\'eel vectors in the two leads are parallel $\bm{n}_0 = \bm{n}_1$. In Fig. \ref{fig:NLGvsEV}(a), we plot the normalized nonlocal conductance $\mathcal{G}_{\mathrm{NL}} = G_{\mathrm{NL}}/\sum_s G_1^s$ at zero temperature as a function of the voltage bias $eV/\Delta_0$ for the AF-S-AF junction.
If the applied voltage is less than the superconducting gap, that is, $eV/\Delta_0<1$, both the CT and AR signals are completely suppressed due to the antiferromagnetic exchange gap. In this regime, the nonlocal conductance is negative, and thus, the CAR signal is dominant. The amplitude of the nonlocal conductance depends strongly on $L_{\mathrm{S}}/\xi$; see the SM \cite{supplement}.

Thus far, we have considered the N\'eel vectors to be parallel. We show in the inset in Fig. \ref{fig:NLGvsEV}(a) that the amplitude of the total nonlocal conductance varies with the misalignment angle between the two N\'eel vectors, while its sign remains unchanged.
We attribute this anisotropic CAR signal to the opening of spin-flip channels during the scattering processes.

For completeness, we compare our result for 2D antiferromagnetic hexagonal leads, as shown in Fig. \ref{fig:NLGvsEV}(a), with those of nonmagnetic graphene and 2D ferromagnetic hexagonal leads, which have previously been reported in the literature \cite{PhysRevLett.100.147001,PhysRevB.80.014513}.

In Fig. \ref{fig:NLGvsEV}(b), we plot the nonlocal conductance of an NM-S-NM heterostructure, where NM is a graphene layer, by setting $J = 0$ in the BdG Hamiltonian \eqref{Eq:DBDG}. In this case, AR and CT are completely suppressed only at $eV = E_{\mathrm{F}}<\Delta_0$ \cite{PhysRevLett.100.147001}. For other bias voltages, competition among the AR, CT, and CAR signals occurs, which, for certain parameters, can lead to a negative nonlocal conductance, as shown in Fig. \ref{fig:NLGvsEV}(b). In contrast, the nonlocal conductance in the AF-S-AF junction is negative for all subgap voltages under the conditions $E_{\mathrm{F}} = J$ and $E_{\mathrm{F}}/\Delta_0>1/2$.
In the NM-S-NM junction, the CAR dominant signal is predicted only when the local Fermi energy is smaller than the superconducting gap. In this regime, inevitable spatial fluctuations in the carrier density, and consequently, the local Fermi energy, in graphene layers are larger than the superconducting gap and hinder experimental detection of CAR signals \cite{Efetov2016,Park2019}. In sharp contrast, in the AF-S-AF junction, a CAR-dominant signal can be observed when the local Fermi energy is larger than the superconducting gap. We therefore expect experimental detection of the CAR-dominant signal to be considerably easier in AF-S-AF junctions than in 2D NM-S-NM junctions.

In 2D hexagonal FM-S-FM junctions, when the magnetization vectors in the two leads are parallel (antiparallel) and both leads have the same charge doping, the CT (CAR) signal dominates the total nonlocal conductance only if the ferromagnetic exchange energy is equal to the local Fermi energy and much larger than the superconducting gap \cite{PhysRevB.80.014513}. These features are demonstrated in Fig. \ref{fig:NLGvsEV}(c), where we plot the nonlocal conductance of a 2D FM-S-FM junction in the antiparallel configuration. As shown in the SM \cite {supplement}, the sign of the total nonlocal conductance in FM-S-FM junctions is very sensitive to the angle between two magnetization vectors in the leads, in contrast to the robustness of the sign of the total nonlocal conductance in the antiferromagnetic case. We also emphasize that in 2D hexagonal FM-S-FM junctions, a CAR-dominant regime is only achieved when the exchange interaction is fine-tuned to the Fermi energy of the ferromagnetic lead. However, in this regime, the density of states for minority spins is negligible, and thus, the electron spins in the two ferromagnetic leads cannot be totally entangled. In the antiferromagnetic leads, the spins are degenerate and truly spin-entangled particles can be generated in two spatially separated leads.

\begin{figure}[t]
\centering
\includegraphics[width=\columnwidth]{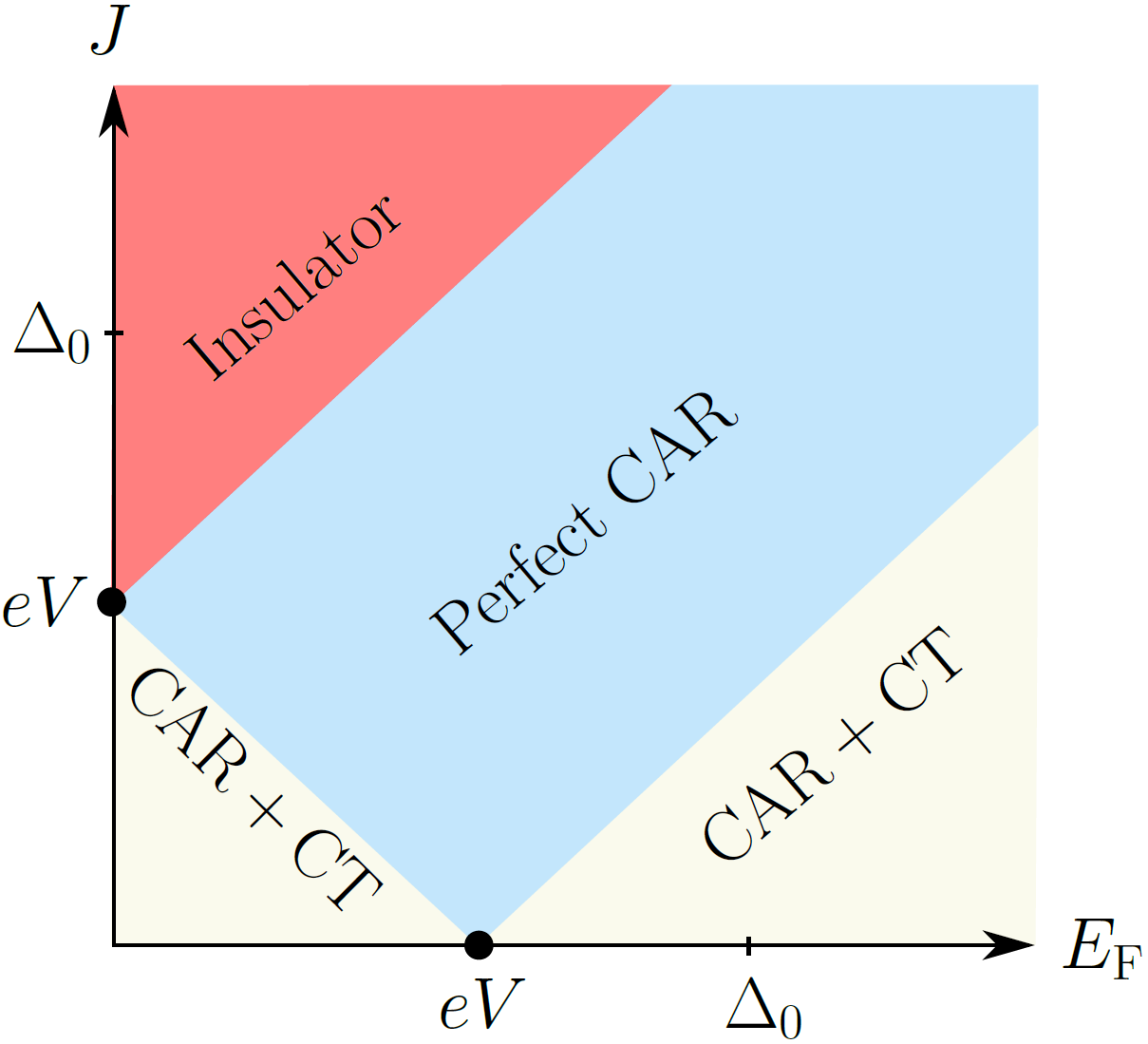}
\caption{Plot of the parameter space $\left(E_{\mathrm{F}},J\right)$ of the total nonlocal conductance.
In the blue region, the CAR signal dominates. In the beige regions, the CAR signal competes with CT. In the red region, the antiferromagnetic leads are insulating, and the conductance vanishes. $J=0$ and $E_{\mathrm{F}} = 0$ represent limits in which the leads are nonmagnetic graphene and undoped AFs, respectively.}
\label{fig:robustnesssketch}
\end{figure}

Finally, we comment on the CAR-dominated signal in the AF-S-AF junction when we relax the conditions $E_{\mathrm{F}} = J$ and $E_{\mathrm{F}}/\Delta_0>1/2$ but still maintain $J_0 = J_1=J$ such that both AR and CT are simultaneously nonzero. Figure \ref{fig:robustnesssketch} shows a sketch of the regions in the parameter space $(E_{\mathrm{F}},J)$, where CT and CAR contribute to the nonlocal conductance in the subgap regime $eV/\Delta_0<1$. We can achieve perfect CAR if the deviation of the gate-controlled local Fermi energy $E_{\mathrm{F}}$ from the antiferromagnetic exchange energy $J$ is smaller than the voltage bias, as shown in the light blue region. For larger deviations, as shown in the beige region, competition between CT and CAR determines the sign of the nonlocal conductance. Gradually, CT dominates, resulting in positive nonlocal conductance for large deviations. If the exchange interaction is significantly larger than both the local Fermi energy and the voltage bias, then the system behaves as an insulator with zero conductance, as demonstrated by the red region. In the SM \cite{supplement}, we plot the nonlocal conductance for specific material parameters, demonstrating the general behavior shown in Fig. \ref{fig:robustnesssketch}.

%%%%%%%%%%%%%%%%%%%%%%%%%%%%%%%%%%%%%%%%%%%%%%%%%%%%%%%

\textit{Concluding remarks.}-- We develop a general framework for nonlocal transport in a 2D AF-S-AF heterostructure. Perfect CAR is possible using a gate voltage to tune the local Fermi energy close to the exchange strength, while the two antiferromagnetic leads have opposite charge doping. 
Our finding is quite generic for an important class of collinear two-sublattice AF materials with either hexagonal or square lattice structure.
We propose a concrete experimental requirement: the local Fermi energy deviation from the antiferromagnetic exchange strength should be smaller than the voltage bias. Typical values for the \textit{s-d} exchange interaction can vary from $\mathrm{meV}$ to $\mathrm{eV}$ \cite{PhysRevB.95.014403, PhysRevB.77.115406}, the superconducting gap is typically on the $\mathrm{meV}$ scale \cite{PhysRevLett.126.147701, Park2019}, and the Fermi energy can be tuned by a gate voltage. Hence, 2D antiferromagnetic-based heterostructures exhibit highly electrically controllable Cooper pair splitting in a spin-degenerate system and enable the production of truly entangled electron pairs in solid-state quantum entanglement devices.

%%%%%%%%%%%%%%%%%%%%%%%%%%%%%%%%%%%%%%%%%%%%%%%%%%%
\acknowledgments
This research was supported by the Research Council of Norway through its Centres of Excellence funding scheme (Project No. $262633$, ``QuSpin'') and the Norwegian Financial Mechanism 2014-2021 Project No. 2019/34/H/ST3/00515 (``2Dtronics'').

%%%%%%%%%%%%%%%%%%%%%%%%%%%%%%%%%%%%%%%%%%%%%%%%%%%
%\bibliography{references}

%%%%%%%%%%%%%%%%%%%%%%%%%%%%%%%%%%%%%%%%%%%%%%%%%%%

%%%%%%%%%%%%%%%%%%%%%%%%%%%%%%%%%%%%%%%%%%%%%%%%%%%
\end{document}

\title{Supplemental material:\\ Electrically controlled crossed Andreev reflection in two-dimensional antiferromagnets}

\author{Martin F.~Jakobsen}
\affiliation{Center for Quantum Spintronics, Department of Physics, Norwegian University of Science and Technology, NO-7491 Trondheim, Norway}
\author{Arne Brataas}
\affiliation{Center for Quantum Spintronics, Department of Physics, Norwegian University of Science and Technology, NO-7491 Trondheim, Norway}
\author{Alireza Qaiumzadeh}
\affiliation{Center for Quantum Spintronics, Department of Physics, Norwegian University of Science and Technology, NO-7491 Trondheim, Norway}

\maketitle
\section{Scattering processes and Snell's law}
The local and nonlocal conductance in the main text is determined by the possible scattering processes available in the junction. For brevity, we here consider only the antiferromagnet--superconductor--antiferromagnet (AF--S--AF) junction where the spin is degenerate. The left (right) AF will be referred to as lead $0$ $(1)$.

Consider an incident electron $(e)$ with angle of incidence $\theta$ in lead 0. The possible scattering processes are:
\begin{enumerate}
    \item Reflected electron into lead $0$ (NR) with angle of reflection $\theta_e^0$.
    \item Reflected hole into lead $0$ (AR) with angle of reflection $\theta_h^0$.
    \item Transmitted electron into lead $1$ (CT) with angle of transmission $\theta_e^1$.
    \item Transmitted hole into lead $1$ (CAR) with angle of transmission $\theta_h^1$.
\end{enumerate}
Due to the translational invariance in the $y$-direction we can write down Snell's law for the possible scattering processes,
\begin{equation}
    \begin{split}
        k_e^0 \sin \theta = k_e^0 \sin \theta_e^0 = k_h^0 \sin \theta_h^0 =k_e^1 \sin \theta_e^1 = k_h^1 \sin \theta_h^1.
    \end{split}
    \label{eq:Snells_Law}
\end{equation}
Here the wavenumbers are
\begin{equation}
    \hbar v_{\mathrm{F}}k^{j}_{e(h)} = \sqrt{\left(E_{\mathrm{F}j} \pm E\right)^2 - J_j^2},
    \label{eq:electron_dispersion}
\end{equation}
and the index $j=\{0,1\}$ labels the lead. Equations \eqref{eq:Snells_Law} and \eqref{eq:electron_dispersion} are used to determine the angles of reflection $\theta_{e(h)}^0$ and transmission $\theta_{e(h)}^1$, for a given angle of incidence $\theta$. Furthermore, the scattering processes AR, CT, and CAR have the corresponding critical angles 
\begin{equation}
    \begin{split}
        &\theta_{c,h}^0=\arcsin{\frac{k_h^0}{k_e^0} }, \quad \theta_{c,e}^1=\arcsin{\frac{k_e^1}{k_e^0} },\quad\, \mathrm{and}\quad \theta_{c,h}^1=\arcsin{\frac{k_h^1}{k_e^0} },
    \end{split}
\end{equation}
respectively. For angles of incidence $\theta$ greater (smaller) than a given critical angle the corresponding scattering processes is an evanescent (propagating) state, giving zero (finite) contribution to the relevant conductance.

%%%%%%%%%%%%%%%%%%%%%%%%%%%%%%%%%%%%%%%%%%%%%%%%%%%
\section{Solution Ansatz of the BdG equation}
We can express the wavefunctions in lead $0$, the superconductor, and lead $1$ as
\begin{equation}
\begin{split}
    &\Psi_0 = \Psi_{\mathrm{In}} + \sum_{s=\{\mathrm{P},\mathrm{AP}\}}\left( r_{e,s} \Psi_{r_{e,s}} + r_{h,s} \Psi_{r_{h,s}}\right),\\
    &\Psi_{\mathrm{S}}=\sum_{i=1}^8 a_i \Phi_i,\\
    &\Psi_1 = \sum_{s=\{\mathrm{P},\mathrm{AP}\}}\left( t_{e,s} \Psi_{t_{e,s}} + t_{h,s} \Psi_{t_{h,s}}\right),\\
\end{split}
\end{equation}
respectively. Here $\Psi_{r_{e(h),s}}$ and $\Psi_{t_{e(h),s}}$ are the eigenstates corresponding to the scattering processes NR (AR) and CT (CAR) in the leads. The eigenstates in the superconductor is denoted by $\Phi_i$. The spin-index $s = \mathrm{(A)P}$ refers to the spin polarization being (anti)parallel to the N\'eel vector. The reflection $r_{e(h),s}$ and transmission $t_{e(h),s}$ amplitudes are then determined by requiring continuity of the wavefunction
\begin{equation}
    \Psi_0\left(x=-L_{\mathrm{S}}/2\right)=\Psi_S\left(x=-L_{\mathrm{S}}/2\right)\, \mathrm{and}\, \Psi_S\left(x=L_{\mathrm{S}}/2\right)=\Psi_1\left(x=L_{\mathrm{S}}/2\right)
\end{equation}
at the AF--S interfaces.

\section{Robustness of the CAR dominated signal}
In the main text we argued that a perfect CAR dominated signal can be obtained on the voltage interval $\left(0, \Delta_0\right)$ if the gate voltage is tuned such that $E_{\mathrm{F}} = J$, and $E_{\mathrm{F}}/\Delta_0>1/2$. We also discussed the robustness of deviations from these assumptions. If the deviation of $E_{\mathrm{F}}$ from $J$ is greater than the voltage bias then CT dominates the sign of the nonlocal conductance. In this section, we quantitatively demonstrate the competition between CT and CAR.

\begin{figure}[h]
\centering
\includegraphics[width=0.7\columnwidth]{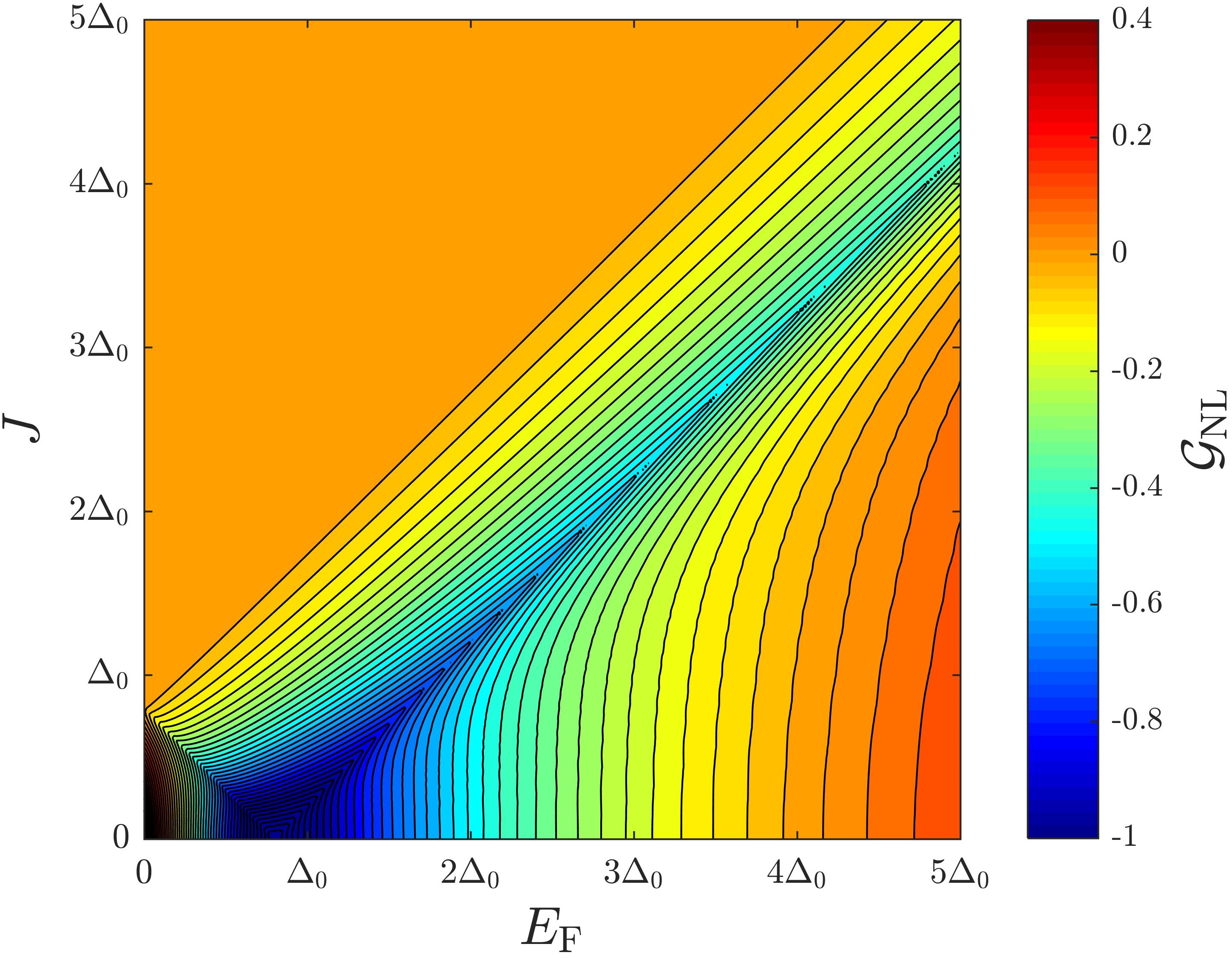}
\caption{Nonlocal conductance as a function of $J$ and $E_{\mathrm{F}}$. The figure is an example of the general behaviour discussed in Fig. 4 of the main text. We have used the concrete parameters $L_{\mathrm{S}}/\xi = 1$ and $eV/\Delta_0 = 0.8$.}
\label{fig:robust}
\end{figure}

In Fig. \ref{fig:robust} we have plotted the nonlocal conductance as a function of the parameters $J$ and $E_{\mathrm{F}}$. Fig. \ref{fig:robust} exhibits the general behaviour of Fig. 4 in the main text. The system is insulating when $J>eV + E_{\mathrm{F}}$, and exhibits perfect CAR when $J> \pm \left( eV-E_{\mathrm{F}} \right)$. When these inequalities are not satisfied there is a competition between CAR and CT, that determines the sign of the nonlocal conductance.

\begin{figure}[h]
\centering
\includegraphics[width=\columnwidth]{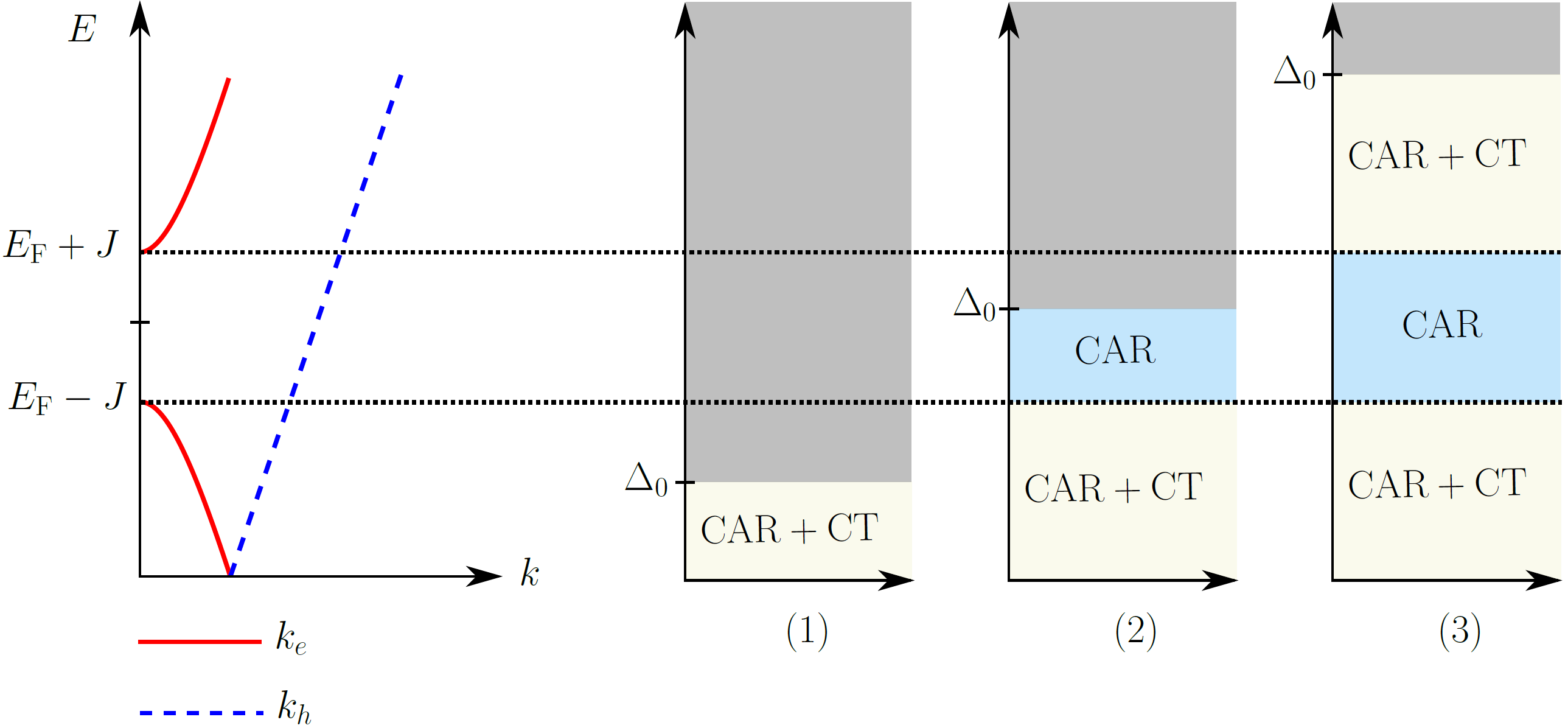}
\caption{The possible scattering processes allowed in lead 1, depending on the relative values of $J$, $E_{\mathrm{F}}$, and $\Delta_0$.}
\label{fig:banddiagramsupplement.png}
\end{figure}

We can understand the general features of Fig. \ref{fig:robust} analytically by considering the band structures of lead 1 and the superconductor in the subgap regime $eV/\Delta_0<1$ shown in Fig. \ref{fig:banddiagramsupplement.png}. There are in total four cases to consider:
\begin{enumerate}
    \item For $\Delta_0 < E_{\mathrm{F}} - J$, CT and CAR competes for all $eV<\Delta_0$.
    \item For $ E_{\mathrm{F}} - J<\Delta_0 < E_{\mathrm{F}} + J$, we find perfect CAR if $E_{\mathrm{F}}- J<eV < \Delta_0$, and competition between CT and CAR if $eV < E_{\mathrm{F}} - J$.
    \item For $\Delta_0 > E_{\mathrm{F}} + J$, we find perfect CAR if $E_{\mathrm{F}}- J < eV < E_{\mathrm{F}}+ J$, and competition between CT and CAR on the intervals $eV < E_{\mathrm{F}}- J$ and $E_{\mathrm{F}}+ J<eV<\Delta_0$.
    \item For $E_{\mathrm{F}} < J$ and $eV < J - E_{\mathrm{F}}$ the system is insulating.
\end{enumerate}
By solving these inequalities in terms of $J$ we obtain Fig. 4 in the main text.

\begin{figure}[h]
\centering
\includegraphics[width=\columnwidth]{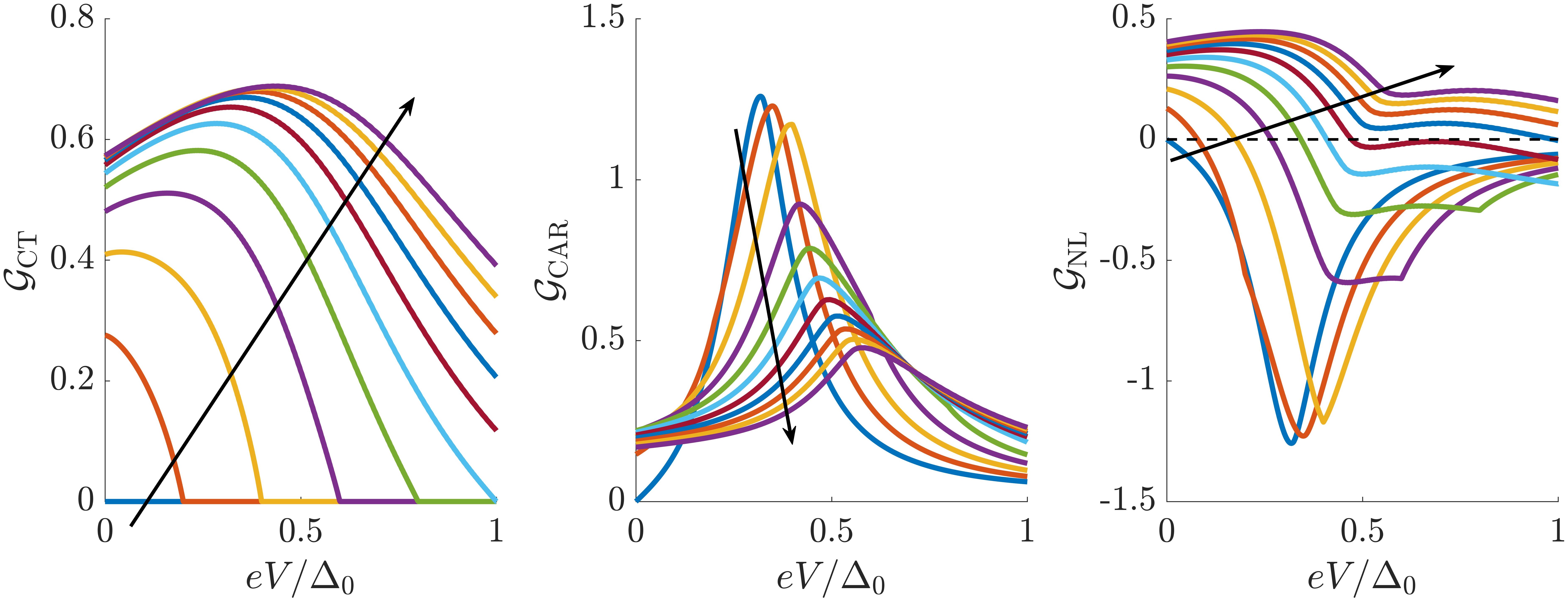}
\caption{A parameter sweep over the deviation parameter $l$. In the figure we have used the values $l = \{0, 0.2, 0.4, 0.6, 0.8, 1, 1.2, 1.4, 1.6, 1.8, 2\}$, and fixed $E_{\mathrm{F}}= 10\Delta_0$. The arrows indicate curves with increasing deviation parameter $l$. As the deviation parameter increases CT starts to gradually dominate CAR. Other choices of $E_{\mathrm{F}}$ result in the same qualitative behaviour.}
\label{fig:deviationsweep}
\end{figure}

In Fig. \ref{fig:deviationsweep} we have plotted the contributions from CT, CAR, and the total nonlocal conductance $\mathcal{G}_{\mathrm{NL}}$ as a function of $eV/\Delta_0$. We have parameterized the exchange interaction as
\begin{equation}
    J = E_{\mathrm{F}} - l\,\Delta_0.
\end{equation}
Here $l$ is a numerical parameter that determines the deviation between $E_{\mathrm{F}}$ and $J$ in terms of $\Delta_0$. In Fig. \ref{fig:deviationsweep} we have performed a parameter sweep over $l$. Concretely we have let $l$ be on the interval $\left(0,2\right)$. The figure shows that as the deviation parameter increases, CT gradually starts to dominate CAR resulting in a sign change of the nonlocal conductance.

\newpage
\section{Oscillations in the Nonlocal conductance}
\begin{figure}[h]
\centering
\includegraphics[width=0.5\columnwidth]{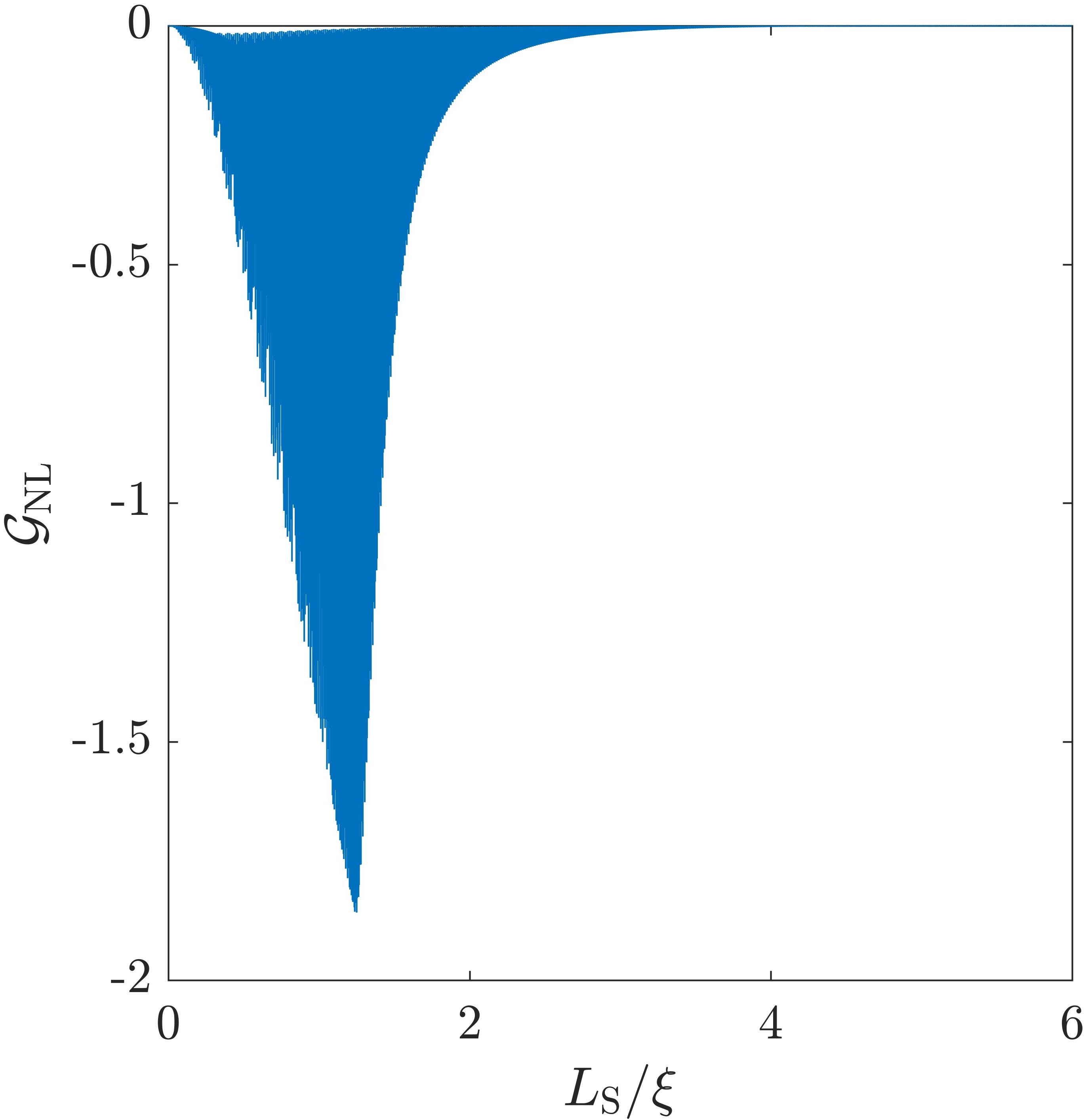}
\caption{Normalized nonlocal conductance  $\mathcal{G}_{\mathrm{NL}}$ as a function of the length of the superconductor $L_{\mathrm{S}}/\xi$. Here $\xi=\hbar v_{\mathrm{F}}/\Delta$ is the approximate length of a Cooper pair. We have set $eV/\Delta_0 = 0.5$, $E_{\mathrm{F}}/\Delta_0 = J/\Delta_0 = 10$, and $E_{\mathrm{FS}}/\Delta_0 = 1000$.}
\label{fig:oscillationls.jpg}
\end{figure}

In Fig. \ref{fig:oscillationls.jpg} we have plotted the nonlocal conductance $\mathcal{G}_{\mathrm{NL}}$ as a function of the length of the superconductor $L_{\mathrm{S}}/\xi$, for $eV/\Delta_0 = 0.5$. We observe rapid oscillations, due to the formation of resonant transmission levels inside the superconductor. The oscillation frequency is determined by the ratio $\Delta_0/E_{\mathrm{FS}}$. The oscillating behaviour is also present in NM--S--NM and FM--S--FM junctions, as confirmed by previous studies referenced in the main text. Importantly, CT is suppressed for all lengths $L_{\mathrm{S}}/\xi$, when $E_{\mathrm{F}} = J$ and $E_{\mathrm{F}}/\Delta_0>1/2$, resulting in a CAR dominant signal. The nonlocal conductance is exponentially suppressed in the limit $L_{\mathrm{S}}/\xi \rightarrow \infty$, where the spatial separation of the electrons in lead 0 and lead 1 is greater than the approximate length of a corresponding Cooper pair.

\newpage
\section{Anisotropy in conductance}
\begin{figure}[h]
\centering
\includegraphics[width=\columnwidth]{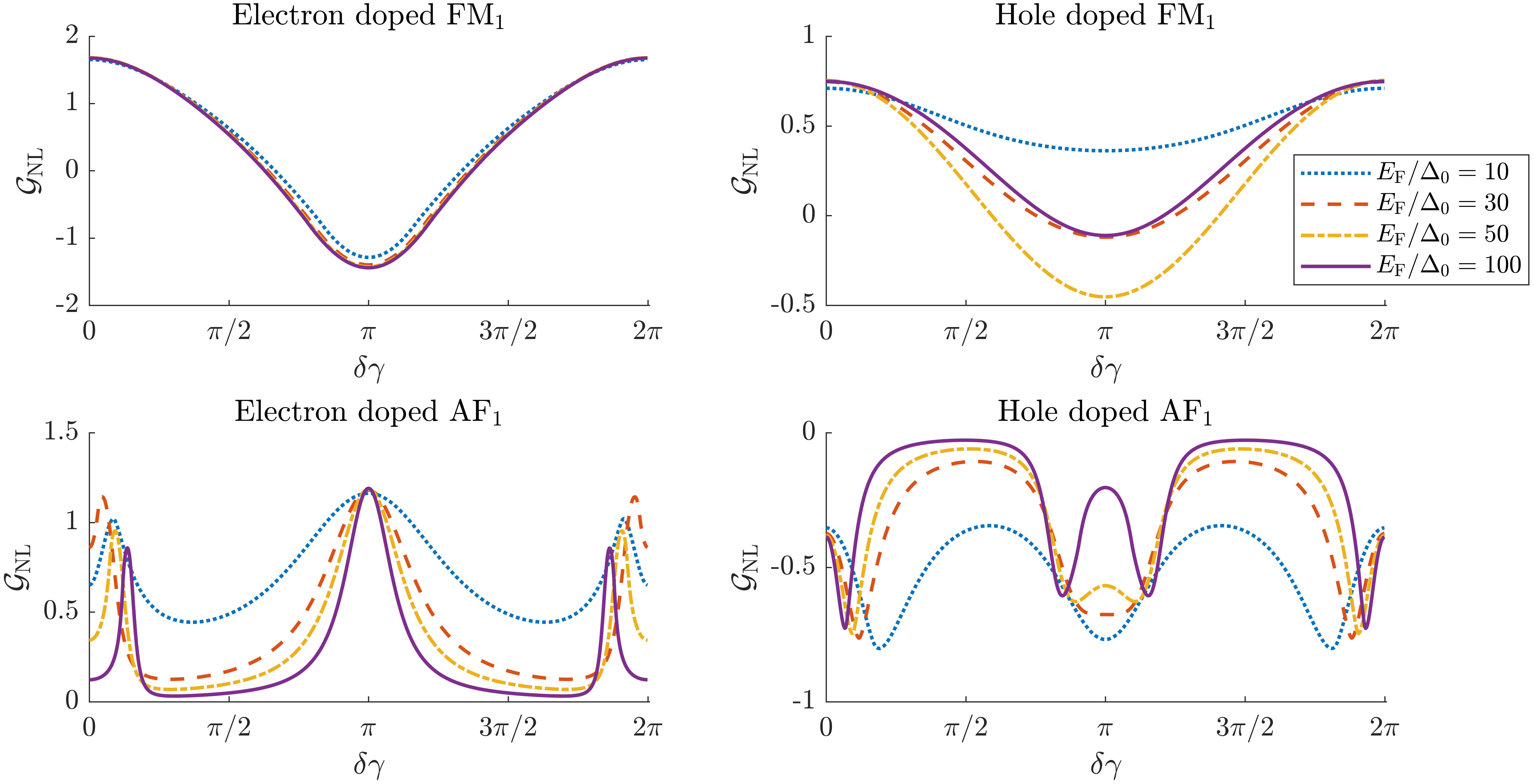}
\caption{Normalized nonlocal conductance $\mathcal{G}_{\mathrm{NL}}$ as a function of the misalignment angle $\delta \gamma$ between the order parameter vectors. In all cases lead $0$ is electron doped. These plots clearly show that the sign of the total nonlocal conductance is very sensitive to the order parameter direction in FM leads, while it remains unchanged in AF leads.}
\label{fig:anisotropy}
\end{figure}

In Fig. \ref{fig:anisotropy} we have plotted the normalized nonlocal conductance $\mathcal{G}_{\mathrm{NL}}$ as a function of the angle of misalignment $\delta \gamma = \arccos{\left(\bm{n}_0\cdot\bm{n}_1\right)}$ between the order parameters $\bm{n}_0$ and $\bm{n}_1$ in lead 0 and lead 1 respectively. We have considered the cases where lead 1 can be: electron (hole) doped (anti)ferromagnets. We have fixed $J = E_{\mathrm{F}} = E_{\mathrm{F}0} = - E_{\mathrm{F}1}$, $L_{\mathrm{S}}/\xi = 1$, and $eV/\Delta_0 = 0.5$. In accordance with the literature mentioned in the main text we find a sign-change effect for the electron doped $\mathrm{FM}_1$. Since, the energy bands are spin-degenerate in $\mathrm{AF}_1$ no sign-change effect is observed. However, all magnetic junctions exhibit an anisotropy as a consequence of the misalignment between the order parameters in lead 0 and lead 1. Notice that for the hole doped $\mathrm{AF}_1$ the nonlocal conductance is negative for any misalignment.

\section{Interface effects}
In the main text, we have assumed that the interface is $i)$ ideal and $ii)$ magnetically compensated. We here want to investigate the effect of lifting these assumptions quantitatively. In the BdG Hamiltonian, interface effects can be modelled by a delta-function-like potential. The interface Hamiltonian, incorporating $i)$ and $ii)$ is 
\begin{equation}
    H_\mathrm{I} = 
    \begin{bmatrix}
    H_{V} + H_{h} & 0 \\
    0 & -H_{V} - H_{h}^T
    \end{bmatrix}\delta(x-x_0),
    \label{eq:BdG_interface_term}
\end{equation}
where $x_0$ is the position of the interface.
We model a non--ideal interface with a spin-independent potential barrier of strength $V>0$,
\begin{equation}
    H_{V} = V s_0 \otimes \sigma_0.
\end{equation}
To incorporate the effect of an uncompensated interface, we introduce the following spin-dependent potential  
\begin{equation}
    H_{h} = \frac{h}{2} \left(\mathbf{n}(x)\cdot \mathbf{s}\right)\otimes \left(\sigma_0 + \sigma_z\right),
\end{equation}
modelling a net magnetization with strength $h$. The Pauli matrices $\bm{s}$ and $\bm{\sigma}$ denote spin and sublattice degrees of freedom.

Adding the interface potential, Eq. \eqref{eq:BdG_interface_term}, to the BdG Hamiltonian of the main text, and integrating over the interface at $x_0$, we obtain the following continuity condition
\begin{equation}
    i \hbar v_{\mathrm{F}}\, \tau_z \otimes s_0 \otimes \sigma_x \left[ \psi(x_0^+) - \psi(x_0^-) \right]=
    \left[V \tau_z \otimes s_0 \otimes \sigma_0
    + \frac{h}{2} \tau_4 \otimes \left(\mathbf{n}(x)\cdot \mathbf{s}\right)\otimes \left(\sigma_0 + \sigma_z\right) \right]\psi(x_0)
\end{equation}
where we introduced the Pauli matrices for charge degrees of freedom (electron/hole)  $\bm{\tau}$  and the matrix $\tau_4 = \mathrm{diag}\left(1,-T\right)$, where $T$ represents the transpose operator.

\begin{figure}[h]
\centering
\includegraphics[width=1\columnwidth]{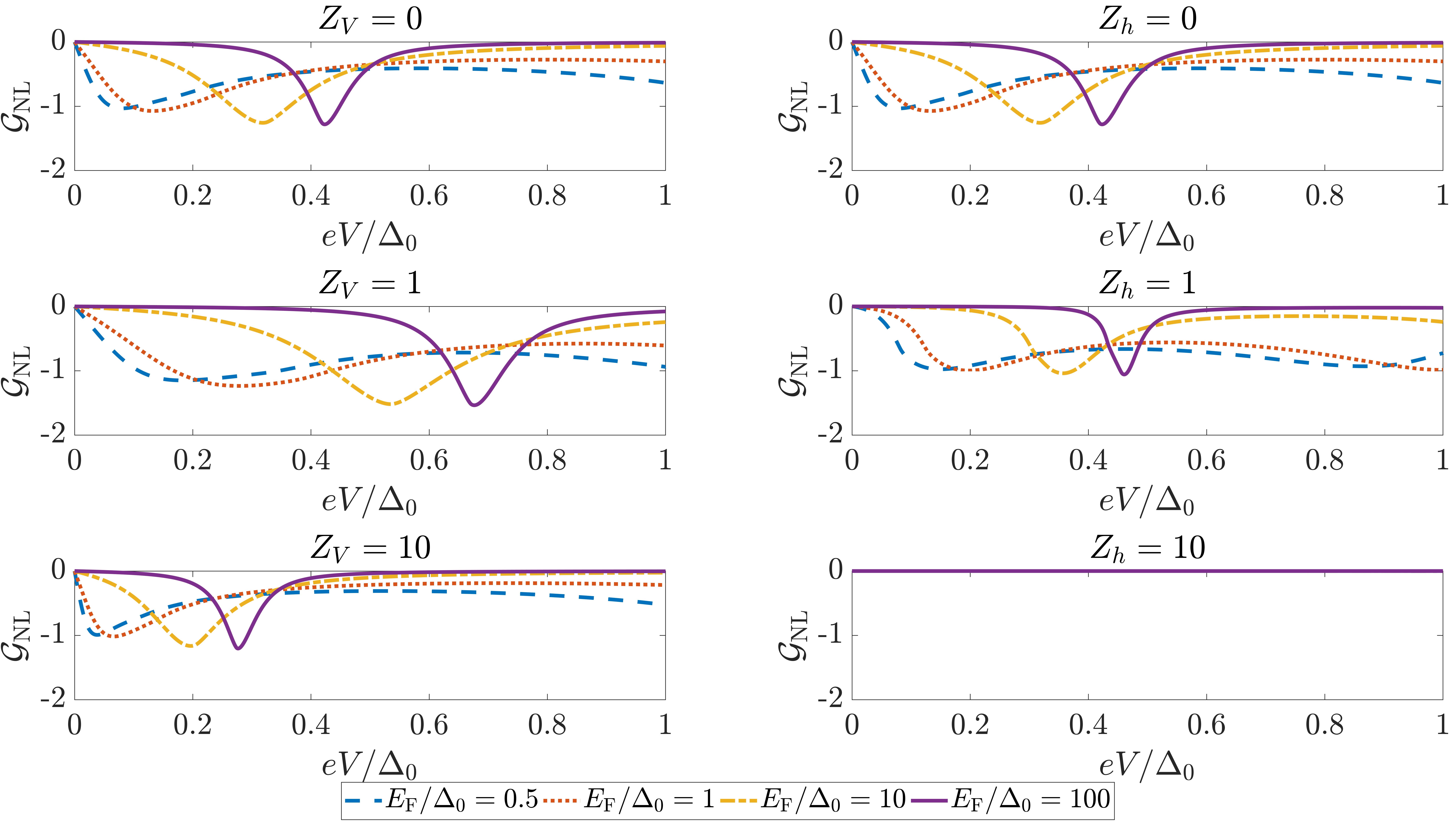}
\caption{Normalized nonlocal conductance $\mathcal{G}_{\mathrm{NL}}$ as a function of the potential barriers $Z_V$ and $Z_h$ parameterizing the transparency of the interface and degree of uncompensated magnetic moments respectively.}
\label{fig:fig6sm}
\end{figure}

To quantitatively investigate the effect of imperfect or uncompensated interfaces, we introduce the dimensionless barrier strength $Z_V = V/\hbar v_{\mathrm{F}}$ and the dimensionless uncompensated magnetic moment $Z_h = h/\hbar v_{\mathrm{F}}$ respectively. In Fig. \ref{fig:fig6sm}, we demonstrate how the interface effects changes the non--local conductance.

An imperfect interface, modelled by spin-independent potential, does not suppress or qualitatively change the non--local conductance, only a shift in the position of the conductance peak is observed. We attribute this behaviour to the ''relativistic'' effects associated with Klein-like tunnelling, masking the potential barrier for hexagonal lattices. 
In non-relativistic Hamiltonians, like a square lattice AF system, we expect that the non--local conductance decreases with increasing the barrier strength.

The effect of an uncompensated interface that leads to a net interface magnetization can be more dramatic. For moderate net magnetizations at the interface, the amplitude of the non--local conductance is slightly reduced. For larger net magnetizations at the interface, the conductance gradually decreases and vanishes completely in the limit of very strong net magnetization. This occurs because a strong  fixed-direction magnetization closes the spin-flip process channels in the system. Since finite local and non--local ARs need a spin-flip channel, both are suppressed in the strong magnetization regime. This leaves NR as the dominant scattering process and thus, both local and non--local conductances are suppressed.

